# Comprehensive Structural Characterization of Charged Polymers Involved in Moisture-Driven Direct Air Capture

Gayathri Yogaganeshan (1 and 2), Rui Zhang (3), Raimund Fromme (1 and 2), Sharang Sharang (4), Jamie Ford (5), Douglas M Yates (5), Marlene Velazco Medel (6), Martin Uher (4), Justin Flory (7), Jennifer Wade (8), Petra Fromme (1 and 2)[*]

((1) Biodesign Center for Applied Structural Discovery, ASU, Tempe, USA, (2) School of Molecular Sciences, ASU, Tempe, USA, (3) Eyring Materials Center, ASU, Tempe, USA, (4) TESCAN USA Inc., (5) Singh Center for Nanotechnology, UPENN, Philadelphia, USA (6) Center for Negative Carbon Emissions, ASU, Tempe, USA, (7) Walton Center for Planetary Health, ASU, Tempe, USA, (8) Mechanical Engineering, NAU, Flagstaff, USA)

**Funding statement**
The author(s) disclosed receipt of the following financial support for the research, authorship, and/or publication of this article: This work was supported by the U.S. Department of Energy, Office of Science, Office of Basic Energy Sciences under Award Number DE-SC0023343.

**Conflicts of Interest**
The authors declare no conflicts of interest related to the content of this paper.

**Abstract**
The rise in atmospheric carbon dioxide ($CO_2$) levels has led to urgent calls for effective carbon capture methods, with direct air capture (DAC) emerging as a promising solution. This study focuses on the structural characterization of commercially available alkaline anion-exchanging membrane (AEM) polymers, Fumasep FAA-3 and IRA 900, for use in low-energy, moisture-driven DAC applications. A combination of X-ray diffraction, small and wide-angle X-ray scattering (SAXS/WAXS), atomic force microscopy (AFM), focused ion beam-scanning electron microscopy (FIB-SEM), and transmission electron microscopy (TEM) was employed to explore the structural features of these materials. X-ray scattering analysis revealed molecular ordering and large-scale structural organization in both materials, while humidity-induced changes highlighted the impact of moisture on structural properties. AFM surface analysis further indicated the presence of clustering, porosity, and swelling, which were corroborated by FIB-SEM and TEM imaging. These structural insights offer a deeper understanding of the behavior of AEM-DAC materials during $CO_2$ capture and release, emphasizing the role of moisture in these processes. This work lays the foundation for the development of more energy-efficient DAC polymers, paving the way for improved $CO_2$ capture technologies.

1. Introduction

Atmospheric carbon (C) levels in the form of $CO_2$ have risen dramatically over the past century. This has manifested in global warming and has caused many detrimental effects, such as changing weather patterns and increased droughts.[1] There is a very urgent and dire need to reduce atmospheric $CO_2$ levels to mitigate current and future negative effects on the biosphere. Many C-reduction methods have been implemented, such as reforestation, agricultural and soil management, C-biomineralization, and ocean fertilization, which are remediation methods. DAC is an alternative, promising method that captures $CO_2$ directly from the air.[1,2]

DAC approaches must have a sorbent, which can be solid or liquid, to capture $CO_2$ by contacting it chemically or physically, and a regeneration method.[2–5] Current DAC methods require significant input of energy, either by heat or by the application of pressure, to remove $CO_2$.[2,3,6,7] Our approach is to utilize alkaline AEM polymers, which comprise positively charged cation groups, specifically a quaternary ammonium cation, to enable binding and the mobility of negatively charged ions through the membrane.[4] There have also been studies reported on polymers containing phosphonium cation for the DAC.[8] These polymers can capture and remove $CO_2$ from the air. Moreover, capturing and removing $CO_2$ using AEM-DAC polymers can be done only with a change in moisture activity, which is called the moisture swing (MS), which was first proposed by Lackner.[1,2,6,7,9–14] Wang et al. in their studies showed that IRA 900, an AEM investigated in this study, has a DAC efficiency in terms of $CO_2$ capacity of 1.92 mmolg$^{-1}$ at 20% relative humidity (RH) at 25 °C.[15]

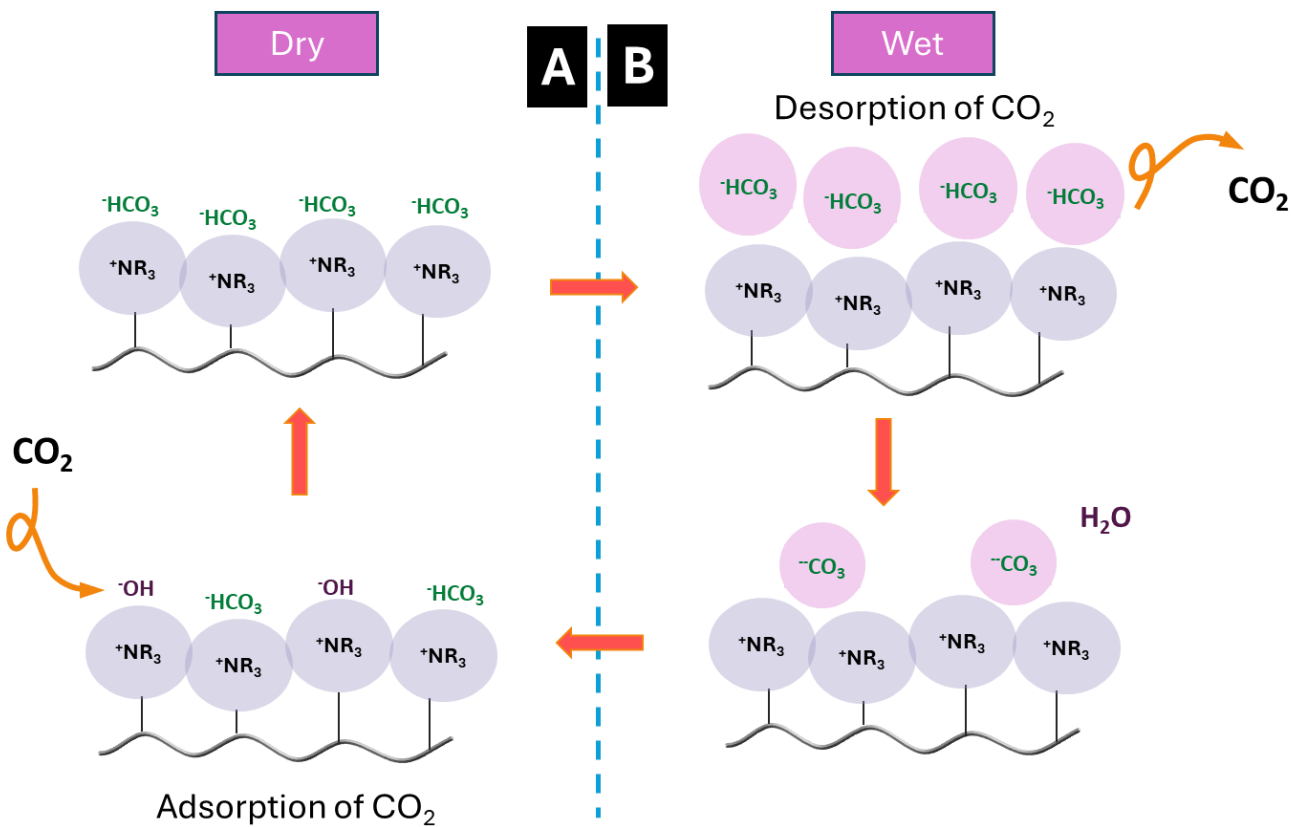

**Figure 1**: Schematics of the MS process. (A) Dry resin. (B) Wet resin.

Figure 1A shows the MS process of a dry, AEM-DAC polymer with quaternary ammonium cations and hydroxide (OH$^-$) counterions that bind $CO_2$ from the air to form bicarbonate ions (HCO$_3^-$) through the reaction, $CO_2 + OH^- \leftrightarrow HCO_3^-$. Figure 1B shows that adding moisture induces two HCO$_3^-$ to form carbonate ions (CO$_3^{2-}$), driving the release of $CO_2$ through the reaction, HCO$_3^- \leftrightarrow$

$CO_2$ + $OH^-$. This is then followed by the moisture-mediated shift to $CO_3^{2-}$ by the reaction, $HCO_3^-$ + $OH^-$ ↔ $CO_3^{2-}$ + $H_2O$. Subsequent drying of the AEM-DAC material in the open air again forms $OH^-$, enabling cyclic $CO_2$ capture. The overall reaction of the $CO_2$ capturing and releasing during the MS cycle is $2HCO_3^-$ ↔ $CO_3^{2-}$ + $CO_2$ + $H_2O$.[1,4,6,7,9–13,15–18]

Even though polymers such as Fumasep, IRA 900, and Marathon A, which can work as DAC polymers, are commercially available, they are brittle, slow to absorb and desorb $CO_2$ with very low capacity, and have yet to be demonstrated with long lifetimes (up to 100,000) needed to be cost-effective.[10] There have been studies on polymers like Fumasep FAA-3 on their mechanical behavior for carbon capture applications.[19] This study is part of Mission DAC, a DOE-funded project that focuses on designing and producing novel DAC polymers that overcome these constraints, which includes structural studies and characterization of the commercially available AEM-DAC polymers as the first essential step.

Using atomic force microscopy studies, Barnes et al. have shown that the structural changes are associated with AEMs while ions exchange from bromide ($Br^-$) to hydroxide form ($OH^-$) in dehydrated and hydrated conditions.[20] Wang et al. have shown the porous nature of these AEMs using SEM.[12] Colletta et al. have shown that a low-dose TEM can be utilized to image the ionomer and its distribution in an AEM.[21] Tian et al. used high-resolution cryo-TEM to study vitrified solutions of nanofibers to combine the structural information with molecular modeling.[22] There was a knowledge gap on the chemical composition and any structural information on commercially available DAC materials in terms of the DAC process. To study the structure of moisture swing materials, techniques such as focused ion beams (FIB) could be used to thin down the sample to a small lamella, but to our knowledge, this technique has not been applied to these materials before. In this work, we aim to determine the structural features of the DAC materials by a combination of AFM, FIB-SEM, TEM, and X-ray scattering methods, such as X-ray diffraction and SAXS/WAXS for structural characterization, where we hypothesize that structured microdomains might exist in these DAC materials. This includes novel technology development with challenges in all aspects, from sample preparation and method development to comprehensive structural studies on such soft polymers.

As Luo et al. mentioned in their work, the hydration levels should play an important role in ion transportation in AEMs.[23] Once the critical structural features influencing the processes of MS, as well as $CO_2$ capture and release, are identified, they can be applied in molecular modeling and polymer synthesis. New polymers that could then remove $CO_2$ from the air more efficiently could be synthesized based on information we obtain on the structure. Here, the relationship between structure and function, structural changes during $CO_2$ adsorption and desorption during the MS process, and structural constraints surrounding the active sites in different and gradient moisture levels will be critical for the design of new DAC materials in the future.

## 2. Materials and methods

### 2.1 Materials
IRA 900 beads (640 – 800 μm in diameter), which have a polystyrene backbone and quaternary ammonium functionality, were purchased from Millipore Sigma. Fumasep FAA-3 30-micrometer films, which have a polyphenylene oxide backbone and quaternary ammonium functionality, were purchased from Fumatech.

### 2.2 Ion exchange

Ion exchange was conducted to activate DAC by submerging the materials (IRA 900/Fumasep FAA-3) twice in 0.5M sodium bicarbonate (from Millipore Sigma) solution (1g of material in 50 mL of solution) for 24 hours.[12,13] Excessive salt was removed from the materials by washing twice with deionized (DI) water. The samples were dried for 24 hours before use.

### 2.3 X-ray diffraction

X-ray diffraction measurements were performed at room temperature using a Rigaku HF007 instrument equipped with a rotating anode with copper (Cu) Kα radiation source, $\lambda = 1.5406$ Å (at 40 kV and 30 mA). The beam size at the interaction point was 70 μm. Ion-exchanged DAC samples were mounted onto an ALS-style base with an 18 mm pin as a sample holder using cyanoacrylate glue. Data collection was done at a 50 mm detector distance. A 1 M Eiger detector (Dectris) was used for data collection. The samples were rotated during X-ray data collection over 180° in the range of -90° to 80° with an oscillation of 0.1° with a step size of 10°/30°, with X-ray data collected for 10-second exposures/ image. Software adxv (2013), a Program to display X-ray Diffraction Images, was utilized for data analysis.

### 2.4 SAXS/WAXS measurements

SAXS/WAXS measurements were performed on a Xenocs Xeuss 3.0 (GI-)SAXS/WAXS/USAXS (ultra small angle X-ray scattering) instrument. A GeniX3D Cu High Flux Very Long (HFVL) focus source was used to produce an 8 keV Cu K alpha collimated X-ray beam with a wavelength of 1.541891 Å (generated at 50 kV and 0.6 mA). A windowless EIGER2 R 1M DECTRIS Hybrid pixel photon counting detector was used to collect the scattering signal.

#### 2.4.1 Measurements under vacuum

The sample-to-detector distances are set at 50 mm, 370 mm, and 900 mm (denoted as WAXS, MAXS, and SAXS in the Xeuss system) to cover a broad Q range between ~0.0035 Å$^{-1}$ and ~3.64 Å$^{-1}$, which corresponds to a length scale range between ~1.73 Å and ~1761.97 Å. A "high resolution" configuration was used for WAXS with a beam size of 0.4 mm, for MAXS with a beam size of 0.4 mm, and SAXS with a beam size of 0.25 mm. A solid stage was used for mounting the ion-exchanged Fumasep FAA-3 film samples in parallel and perpendicular directions. The capillary stage was used for the ion-exchanged IRA 900 beads. The measuring time was 5 mins for WAXS, 7.5 mins for MAXS, and 10 mins for SAXS, with a "line eraser" to double the measuring time for each. Direct beam or empty capillary for the same amount of time was measured for each configuration condition.

#### 2.4.2 SAXS/WAXS under Humidity Conditions

The sample-to-detector distances are set at 64 mm, 370 mm, and 900 mm (denoted as WAXS, MAXS, and SAXS in the Xeuss system) to cover a broad Q range between ~0.0042 Å$^{-1}$ and ~2.93 Å$^{-1}$, which corresponds to a length scale range between ~2.14 Å and ~1480.14 Å. A "high resolution" configuration was used for WAXS with a beam size of 0.4 mm, for MAXS with a beam size of 0.4 mm, and for SAXS with a beam size of 0.25 mm. A humidity stage was used for mounting the ion-exchanged Fumasep FAA-3 film in perpendicular and parallel directions, and IRA 900 beads between two mounted Kapton foils (with holes) at 25 %, 50 %, 75 %, and 95 % humidity conditions. The measuring time was 7.5 mins for WAXS, 10 mins for MAXS, and 15 minutes for SAXS, with a "line eraser" to double the measuring time. An empty cell or two Kapton foil windows for the same amount of time was measured for each configuration condition.

### 2.4.3 SAXS/WAXS data reduction

Each raw 2D scattering image was reduced by azimuthal average (in 360º and 20º sectors in parallel and perpendicular directions on the detector projection plane) to a 1D scattering curve, considering geometrical corrections and transmitted intensity. Pixels near the direct beam and invalid pixels were masked during the data reduction. For each sample, the 1D scattering curve of each configuration was subtracted by the corresponding 1D scattering curve of the direct beam or empty capillary or the Kapton windows, without considering sample thickness, and merged accordingly. Each raw 2D scattering image was also reduced to a 1D scattering curve as a function of azimuthal angle, considering geometrical corrections and transmitted intensity. The range of Q crown is between 0.02 Å$^{-1}$ and 0.1 Å$^{-1}$ for MAXS and SAXS, and between 1 Å$^{-1}$ and 2.3 Å$^{-1}$ for WAXS under vacuum and between 1 Å$^{-1}$ and 1.8 Å$^{-1}$ for WAXS under different humidity conditions. The degree of orientation and Hermans' orientation factor were estimated based on the obtained 1D azimuthal profiles.

## 2.5 AFM
### 2.5.1 AFM imaging in air

AFM was performed using Bruker MultiMode 8 (MM8) and Bruker Dimension Icon AFMs. Experiments in air were conducted by mounting the ion-exchanged DAC samples (Fumasep as film and IRA 900 as bead) on an AFM specimen disc using cyanoacrylate glue (Krazy Glue®). A DNP-10 probe (the A cantilever, tip radius between 20-60 nm with a nominal spring constant of 0.35 N/m)/ Scout150 SS Ral was mounted in the cell. An "optical lever" was used to couple the cantilever motion to the position-sensitive detector, and the cantilever's deflection sensitivity was estimated by the thermal tune method.[24] Both tapping mode and contact mode images were acquired, and topographical images were acquired and saved in Bruker's SPM format. Data analysis was carried out using software from Bruker (Nanoscope Analysis, version 2.0, build R1Sr2.157217) to extract qualitative surface characteristics, including roughness and relative feature dimensions. Multiple scans from different sample regions were conducted to ensure reproducibility.

### 2.5.2 AFM imaging in a liquid cell

To enable imaging of the polymer during and after hydration, a fluid cell was used. The fluid cell components were cleaned with isopropanol and deionized water, dried under clean airflow, and reassembled to ensure leak-free operation. Several probes were used for liquid cell imaging: early experiments were done with Bruker DNP-10 probes (the A cantilever, tip radius between 20-60 nm, and nominal spring constant of 0.35 N/m), with later experiments using ScanAsyst-Liquid+ probes (nominal tip radius and spring constant of 2 nm and 0.70 N/m, respectively). During the liquid cell experiments, glass coverslips were mounted to AFM specimen discs (Electron Microscopy Sciences P/N: 75010) by cyanoacrylate adhesive. The ion-exchanged DAC samples were prepared by mounting them to the coverslips in two ways: by cyanoacrylate glue (Krazy Glue®) for earlier experiments. The strong excess of water in the cell was found to drive additional polymerization of unreacted cyanoacrylates, necessitating the use of an alternative, acrylic adhesive from Adhesive Research. The O-ring, which connects the liquid cell, was then mounted on top of the glass coverslip with cyanoacrylate glue by keeping the sample in the middle; both silicone (Bruker P/N: FCO-10) and Viton (Bruker P/N: VTFCO-10) O-rings were used (determined by local availability). A protective layer of Parafilm® was stretched and placed over the scanner; the sample was then placed on top of the scanner. The fluid cell was placed over the sample such that the O-ring fit into the sealing groove on the bottom surface of the fluid cell. The probe was aligned, and the cantilever's deflection sensitivity was estimated by the thermal tune method, and the cantilever was tuned to its resonance frequency. Both contact mode and tapping mode images were acquired. Analysis was carried out using Bruker (Nanoscope Analysis, version 2.0, build R1Sr2.157217) and Gwyddion 2.67.

### 2.6 FIB-SEM milling and imaging
#### 2.6.1 FIB SEM milling and imaging at room temperature

FIB-SEM was performed using the room temperature TESCAN AMBERX plasma FIB-SEM system. Ion-exchanged Fumasep FAA-3 film, or IRA 900 beads, were mounted as it is on standard SEM stubs using graphite conductive adhesive (from Electron Microscopy Sciences). A 20 nm conductive coating of platinum (Pt) was deposited using a Leica ACE sputter coater. The electron beam parameters were set based on imaging requirements for low-keV surface imaging. Ion beam operations were conducted at 30 kV using Xenon (Xe) ions, with precise calibration of the beam alignment. Regions of interest were defined for milling, etching, or deposition, with milling patterns adjusted for dwell time and pass count. After trenching, the redeposition was cleaned by polishing. High-resolution images were acquired, with data saved in standard formats and analyzed using TESCAN Essence software for feature dimensions. After undercutting, with precise polishing of the lamella to a thickness of < 100 nm, lamellas were attached to the Cu half grids using an ion beam deposited Pt weld.

#### 2.6.2 Cryogenic FIB SEM milling and imaging

For cryogenic (cryo) conditions, a 100% humidified sample was prepared by immersing the material in deionized water overnight. Excess water was carefully removed using Kim wipes before mounting the sample onto a stub with graphite conductive adhesive. A Leica VCM plunger equipped with a cryo arm was used to plunge-freeze the samples. The sample was subsequently transferred to the TESCAN AMBERX Xe plasma FIB-SEM system under cryo conditions, utilizing the cryo shuttle (LeicaVCT) to prevent ice formation. To enhance conductivity, a 20 nm Pt coating was deposited using a Leica ACE sputter coater, with the sample stub handled via the cryo shuttle to maintain temperature stability. Cryo-FIB-SEM was performed on the TESCAN AMBERX Xe plasma FIB-SEM system, operating under Cryo conditions. The electron beam parameters were adjusted to meet the requirements for low-keV-based surface-sensitive imaging. Ion beam operations were conducted at 30 kV using Xe ions, with meticulous calibration of beam alignment. Regions of interest were defined for milling, etching, and deposition. After trenching, surface amorphization was minimized through lower probe current polishing. High-resolution images were acquired, with data saved in standard formats and analyzed using TESCAN Essence software to quantify feature dimensions.

### 2.7 TEM imaging

TEM was carried out using a JEOL F200 operating at an accelerating voltage of 200 kV. The analysis was performed on a thin lamella mounted on Cu grids, which were prepared using a plasma-FIB. The FIB-prepared grids containing the lamella were carefully loaded into the TEM sample holder and inserted into the microscope column. All acquired images were saved in standard .tiff format for subsequent analysis.

## 3 Results and discussion

## 3.1 Structural organization in DAC materials

X-ray diffraction and SAXS/WAXS experiments were conducted on Fumasep FAA-3 films and IRA 900 beads to investigate the structural characteristics of these two DAC materials. Data on the Fumasep FAA-3 films were collected both in parallel and perpendicular orientation to the plane of the film. The results are shown in Figure 2.

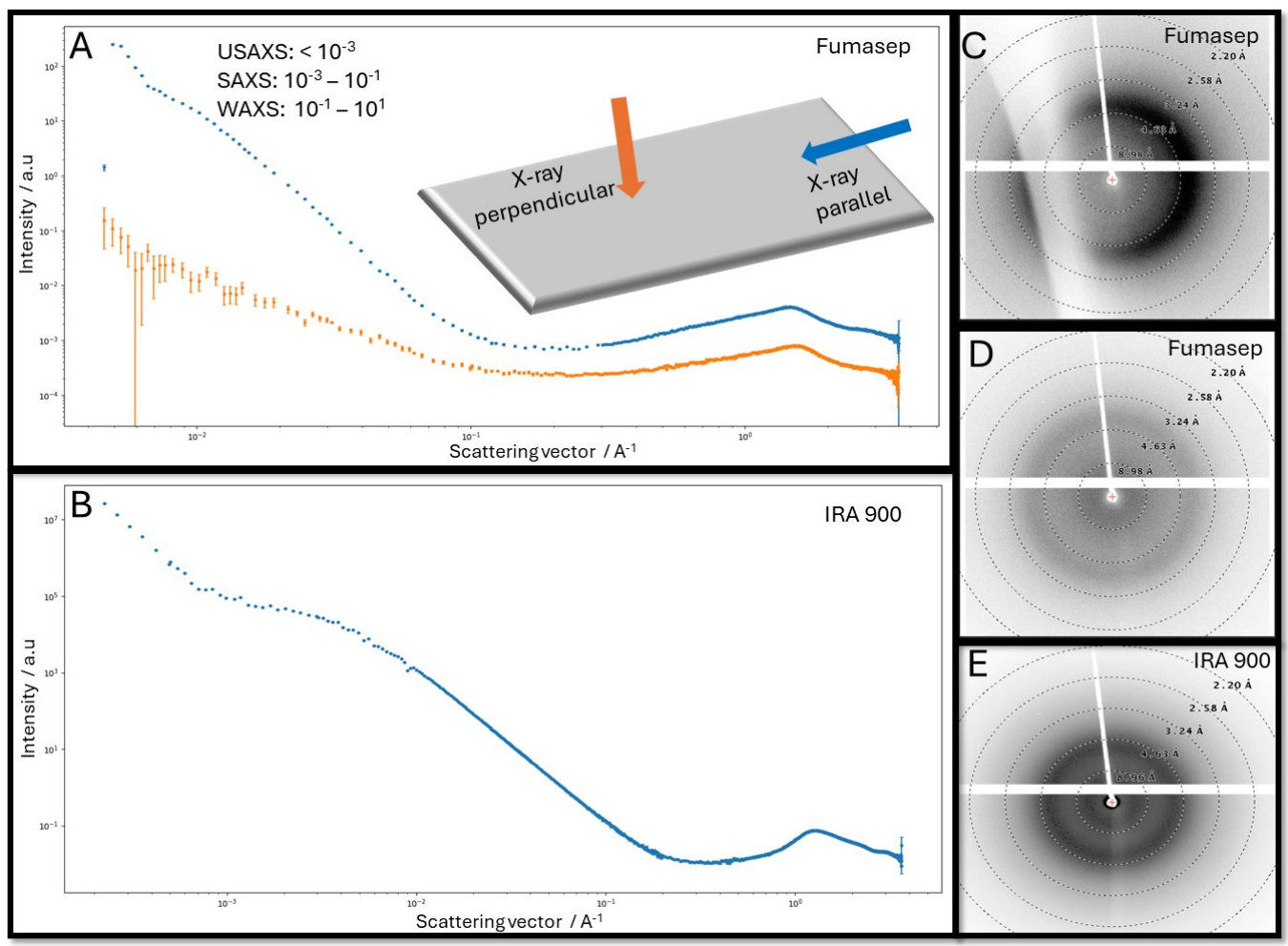

**Figure 2**: Structural organization in DAC materials. (A) Intensity (a.u.) vs scattering vector Q (Å$^{-1}$) plot of 360-degree azimuthal averaged SAXS/WAXS data on Fumasep FAA-3. (Refer to Figure SI-1 for scattering patterns) (B) Intensity (a.u.) vs scattering vector Q (Å$^{-1}$) plot of SAXS/WAXS on IRA 900. (Refer to Figure SI-2 for scattering patterns) (C) X-ray diffraction on Fumasep FAA-3 in a parallel direction. (D) X-ray diffraction on Fumasep FAA-3 in a perpendicular direction. (E) X-ray diffraction on IRA 900.

Figure 2A shows the Intensity (a.u.) vs scattering vector Q (Å$^{-1}$). Here, a distinct peak in the WAXS region around 4 Å (Refer to Figure SI-3 for calculations) was detected for Fumasep FAA-3 in both perpendicular and parallel directions, indicating a degree of molecular order (polymer repeating chain length) on an atomic scale. For the IRA 900 beads, the presence of a peak in the WAXS region around 5 Å (Refer to Figure SI-4 for calculations) in the intensity (a.u.) vs scattering vector Q ($Å^{-1}$) plot similarly suggests atomic structural order (polymer repeating chain length), while the plateaus observed in the SAXS region around 1260 Å (Refer to Figure SI4 for calculations) and, again in the USAXS region, point to larger-scale clustering of polymer chains, as shown in Figure 2B.

Further diffraction patterns from a single crystal diffractometer provide additional insights into the structural orientation of Fumasep FAA-3. The parallel-direction diffraction pattern (Figure 2C) reveals some ordering along the thickness of the material, whereas the perpendicular-direction pattern (Figure 2D) suggests a predominantly amorphous structure, while both show the molecular ordering at around 4 Å that matches the WAXS peak in Figure 2A. Regarding IRA 900, the diffraction pattern (Figure 2E) exhibits a ring around 5 Å, indicative of molecular ordering or homogeneity within the material, which matches the WAXS peak in Figure 2B. These combined results highlight both the molecular order and the larger-scale structural organization present in the tested AEM-DAC materials.

### 3.2 Humidity-associated changes in DAC materials from SAXS/WAXS

Further SAXS/WAXS experiments were conducted to investigate humidity-induced structural changes in DAC materials during the DAC process. Here, we adapted the materials to a specific humidity between 25 % and 95 % in the humidified SAXS chamber for 60 minutes before we collected the SAXS/WAXS patterns. The results are shown in Figure 3.

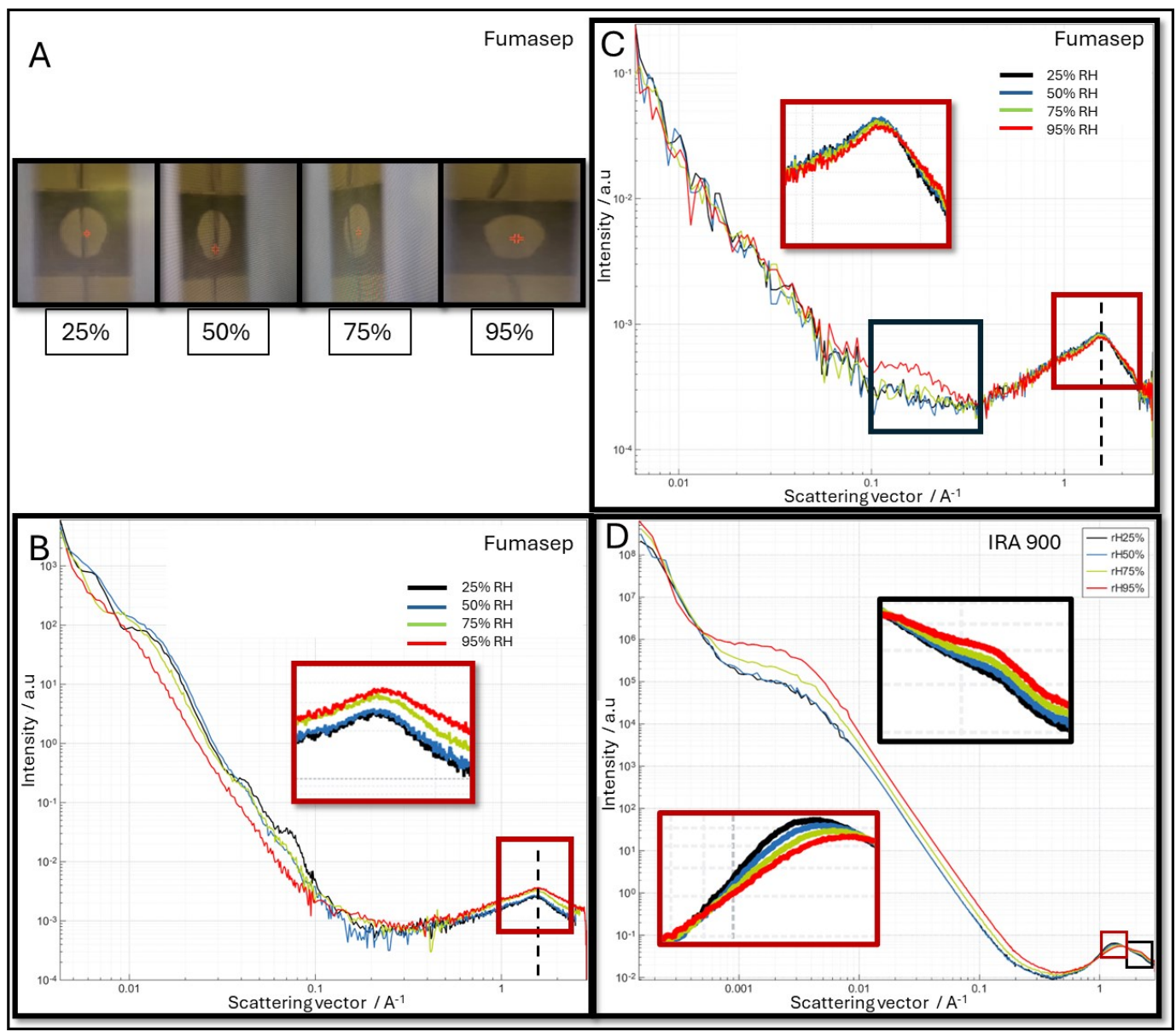

**Figure 3**: Humidity-associated changes in DAC materials. (A) Reversible swelling and shrinking of Fumasep FAA-3 at 25 %, 50 %, 75 %, and 95 % RH in the parallel direction. (B) Intensity (a.u.) vs scattering vector Q (Å$^{-1}$) plot (20-degree azimuthal averaged along horizontal) of SAXS/WAXS of Fumasep FAA-3 at 25 %, 50 %, 75 %, and 95 % RH in parallel direction. (Refer to Figure SI5 for scattering patterns) (C) Intensity (a.u.) vs scattering vector Q (Å$^{-1}$) plot (360-degree azimuthal averaged data) of SAXS/WAXS of Fumasep FAA-3 at 25 %, 50 %, 75 %, and 95 % RH in perpendicular direction. (Refer to Figure SI6 for scattering patterns) (D) Intensity (a.u.) vs scattering vector Q (Å$^{-1}$) plot of SAXS/WAXS of IRA 900 at 25 %, 50 %, 75 %, and 95 % RH. (Refer to Figure SI7 for scattering patterns)

Figure 3A illustrates the bulk swelling behavior of the Fumasep FAA-3 film, in the parallel direction, as RH increases from 25 % to 50 %, 75 % and 95 %. The swelling is reversible as the material swells at higher humidities and shrinks when the RH is reduced to 25 % again. This finding was previously observed for a Fumasep FAA-3 ion exchange membrane. Najibah et al. and Wang et al. showed that when Fumasep FAA-3 membranes are wetted, they swell, and eventual water removal will allow the material to shrink.[15,25] Lopez et al have reported that water uptake in Fumasep FAA-3 at 90% RH and 30 °C is 0.20g/g.[2] Paajanen et al. have studied moisture-induced swelling in wood microfibrils using SAXS/WAXS.[26]

Figures 3B and 3C present the intensity (a.u.) vs scattering vector Q ($Å^{-1}$) curves from the SAXS/WAXS experiments at various humidity levels of 25 %, 50 %, 75 %, and 95 % for Fumasep FAA-3 in the parallel and perpendicular directions, respectively. As humidity increases, the WAXS peak around 4 Å shifts to smaller length scales in both directions (highlighted by a red box in Figures 3B and 3C), suggesting a reduction in the intermolecular distance or polymer building block size or repeating chain length. This finding was initially astonishing, as one would assume that intermolecular distances increase when water is intercalated and the material swells. But as water can't intercalate through a 4Å channel, we speculate that the water inclusion between the repeating units may bring the polymer chains closer together, with tightly packed polymers intercalated by large water channels. In the perpendicular direction, a distinct hump, highlighted by a black box, emerges at 95 % RH around 20 Å to 60 Å (Refer to Figure SI8 for calculations), which might be the appearance of a larger superstructure, potentially due to water uptake and bulk swelling. This is because after 95 % RH, the gaseous water will condense to liquid water, and its inclusion is specifically visible only in the perpendicular direction, but not in the parallel direction.

Figure 3D displays the corresponding SAXS/WAXS results for the IRA 900 beads. The small hump that is around 3 Å (Refer to Figure SI9 for calculations), marked by a black box, increases in intensity, becoming particularly prominent at 95 % RH. The WAXS peak, indicated by a red box that we observed around 5 Å, also shifts to smaller scales with gradually decreasing intensity, suggesting a reduction in intermolecular distances, which may reflect water inclusion bringing the repeating polymer units closer together. Additionally, the SAXS region shows increased plateau intensities, likely pointing to swelling as humidity rises, and we expect the same in the USAXS region.

These results collectively demonstrate how structural changes in Fumasep FAA-3 and IRA 900 materials are influenced by varying humidity levels and evidence that the water uptake plays a major role in ion and water transport in these AEMs, as mentioned by Luo et al., which should also be the same with the DAC process.[23] In their study, the water sorption behavior of AEMs with increasing relative humidity showed a sigmoidal shape change, with the sorption hysteresis maximum around 70 % RH which would attribute to water swelling which would also explain the prominent shift of the small hump with increasing intensity in IRA 900 from 25 % to 95 % RH and as the polymer structure gets condensed with bulk swelling that was visually observable also in Fumasep FAA-3 (Figure 3A).[2,23,27–30]

### 3.3 Surface analysis of DAC materials with AFM

Surface analysis of the Fumasep FAA-3 film and IRA 900 beads were performed using AFM. The goal of these experiments was to detect if domain structures exist in Fumasep FAA-3 and/or IRA 900.

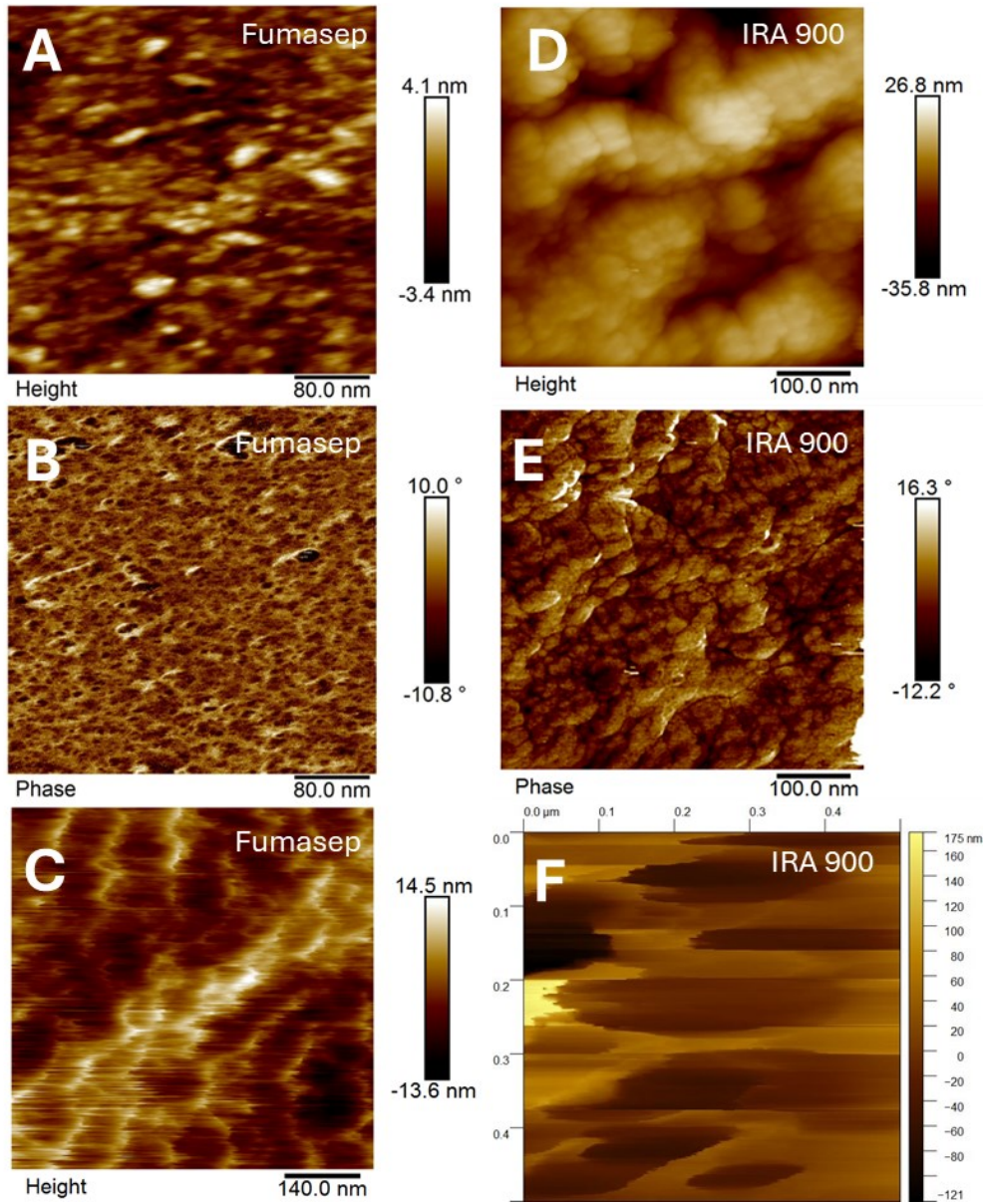

**Figure 4**: Atomic force microscopy of DAC materials. (A) Height image of Fumasep after 0th order, XY plane fit along with (B) the corresponding phase image after 0th-order flattening. (C) Contact mode height image of Fumasep under water after 2nd-order Y plane fit and 0th-order flattening. (D) Height image of IRA 900 after 2nd order XY plane fit, erasing streaks, and using a low-pass filter. (E) Phase image of IRA 900 after 0th order flattening, erasing streaks, and using a low-pass filter. (F) Contact mode height image of IRA 900 underwater after a 2nd order Y background fit, and performing a "median of difference" background fit in Gwyddion.

The height image in Fumasep (Figure 4A) shows contrast from the measured deflection of the cantilever due to the interaction of the tip apex with the sample surface corresponding to the topography of the surface, while Figure 4B displays the corresponding phase image, with contrast resulting from changes in the drive frequency of the cantilever. This "phase contrast" with suitable consideration of the kinds of interactions likely to be present at a given surface, be used to qualitatively infer the mechanical properties of the surface (i.e., how hard or soft the material is at a given point in the image). Figure 4C shows the contact mode height image of Fumasep under water, where the change in surface morphology with respect to the dry image suggests that the hydrated surface undergoes swelling.

For the IRA 900 beads, Figures 4D and 4E show the height and phase images, respectively, revealing clustering at length scales of more than 300 nm and the presence of pores ranging from 70–100 nm. We assume that the clusters that are visible in the AFM image region could resemble the SAXS plateau that we observed in the SAXS/WAXS plot shown in Figure 2B. The nanoporous nature of IRA 900 was also confirmed by Wang et al. and Lopez et al. in their work.[2,15] Figure 4F shows the height image of IRA 900 under water, which suggests swelling and clustering that was also highlighted by Lopez et al. during the water uptake.[2] These AFM results further support the clustering behavior of IRA 900 observed in the SAXS/WAXS experiments, as shown in Figures 2B and 3D. Collectively, these findings provide insight into the nanoscale surface characteristics of these DAC materials, reinforcing the structural insights gained from X-ray scattering measurements.

### 3.4 Structural probing and successful preparation of thin lamellas using FIB-SEM

To obtain a higher resolution structure of our materials, we used Focused Ion Beam Scanning Electron Microscopy (FIB-SEM) to investigate the structure of Fumasep FAA-3 and IRA 900. In FIB-SEM, a focused ion beam is used to mill a sample on the EM grid down to a height of 100-200nm, thereby enabling the electron beam to penetrate and provide an image of the section of the material. While FIB-SEM is widely used in the study of biomolecular structures in their cellular environment, our study is, to our knowledge, the first time FIB-SEM has been applied to soft polymeric anion exchange materials.[31,32] The reason might have been that sample preparation is so much more difficult for these materials compared to the milling of biological materials. For bio samples, cells are grown on the EM-grid and adhere strongly to it, which facilitates the milling process. In contrast, neither the Fumasep FAA-3 membranes nor the IRA 900 beads stick to the grid and fall off as soon as the grid is introduced into the instrument. We had first worked with folding grids, where the Fumasep FAA-3 samples were sandwiched between two metal grids. However, this made the sample grid very thick, which was incompatible with high-resolution imaging in a TEM microscope. Finally, we developed gluing procedures, described in Methods, that could be used to fix the samples on the sample stub. The next challenge was that we determined that the materials were even more radiation sensitive than bio samples, so that milling had to be performed with decreasing current strengths, as we described in the methods. Moreover, the lamellae from these soft materials are very fragile and break very easily when lifted out.

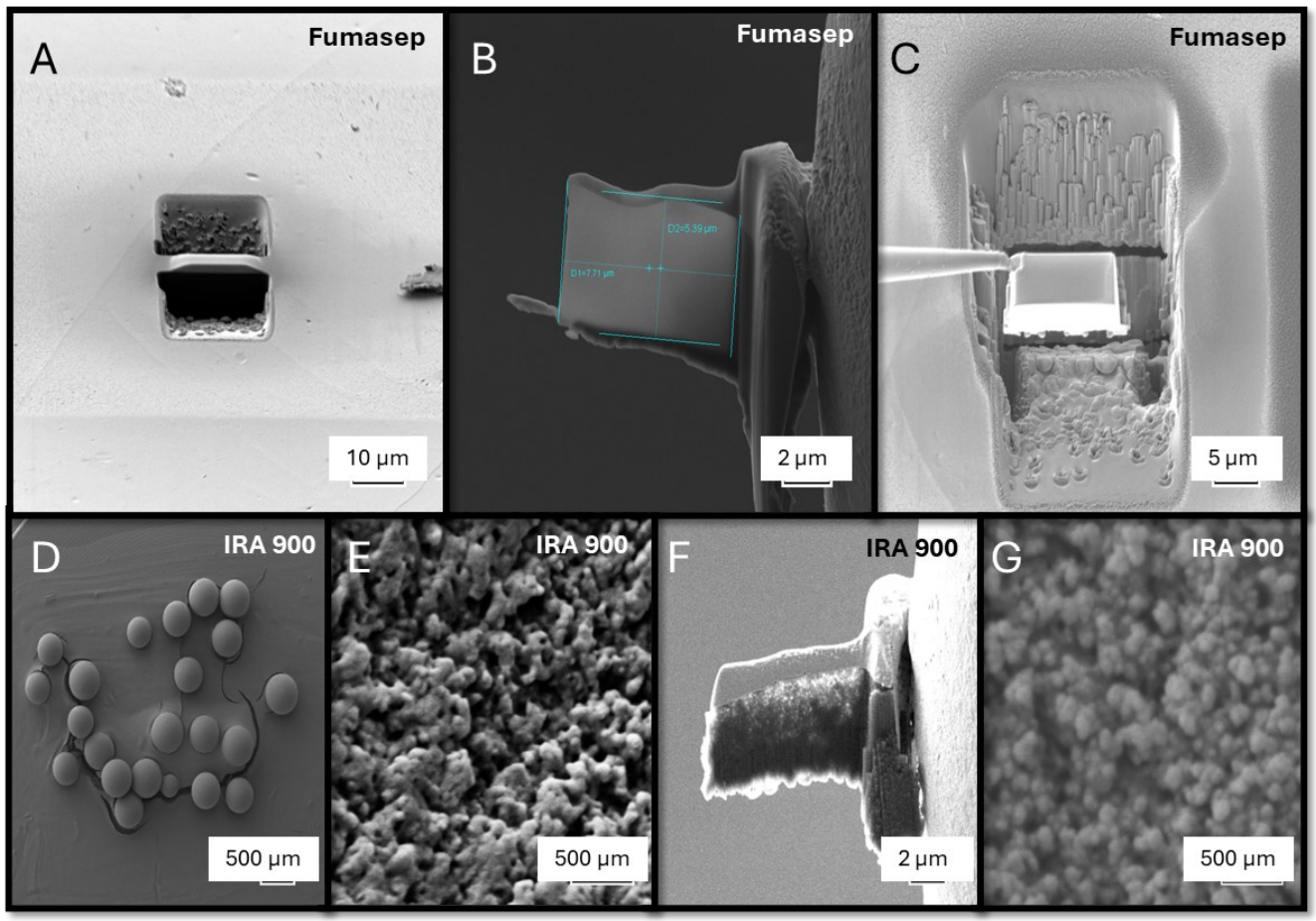

**Figure 5**: Lamella preparation using FIB-SEM in DAC materials. (A) Bridge-type lamella cut in Fumasep FAA-3. (B) Room temperature lamella of Fumasep FAA-3. (C) Lamella lift out in Fumasep FAA-3. (D) SEM image of loaded IRA 900 beads for FIB milling. (E) SEM image of the surface of an IRA 900 bead at ambient (30% RH) conditions. (F) Room temperature lamella of IRA 900. (G) SEM image of the surface of an IRA 900 bead at 100% RH in cryo conditions.

Figure 5 shows the lamella preparation procedure that we established for FIB-SEM experiments. Figures 5A–C depict the successful preparation of a lamella from the Fumasep FAA-3 film at room temperature: 5A shows the bridge-type lamella cut, 5B displays the final prepared lamella, and 5C highlights the successful lamella lift-out. Figure 5D shows an SEM image of IRA 900 immediately after loading into the FIB-SEM system for room-temperature milling. Figure 5E illustrates the surface morphology of IRA 900 captured using the SEM beam, clearly revealing clusters estimated to be around 300 nm and pores around 80 nm, consistent with the AFM results (Figures 4D and 4E). Figure 5F presents the successfully prepared lamella of IRA 900 at room temperature. To investigate the effect of humidity, Figure 5G shows the surface of IRA 900 after immersion in water to achieve 100% RH to prepare the lamella under cryo conditions. Compared to Figure 5E (ambient conditions), Figure 5G reveals increased clustering and reduced porosity at 100% RH. These observations suggest that the clustering of IRA 900 intensifies, and the porous structure reduces with increasing humidity during the DAC process.

### 3.5 TEM images of lamellas showing atomic-scale structural features of DAC materials

The RT lamellas that were successfully prepared using FIB-SEM were subsequently analyzed in TEM to detect higher-resolution structural features. The results are shown in Figure 6.

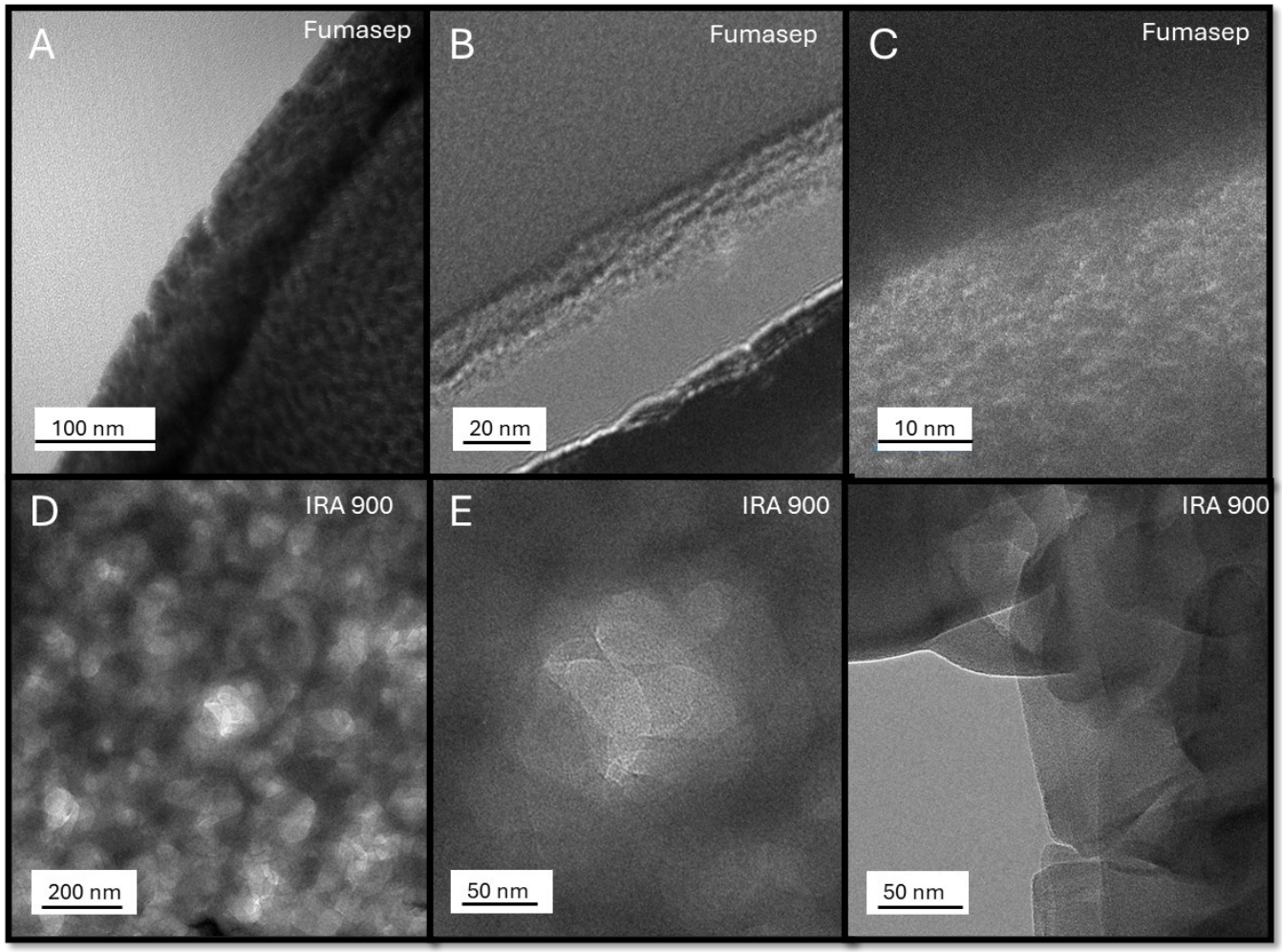

**Figure 6**: TEM imaging of DAC materials. (A) and (B) TEM image of the Room temperature lamella of Fumasep FAA-3 in the parallel direction. (C) TEM image of the Room temperature lamella of Fumasep FAA-3 in the perpendicular direction. (D) TEM image of room-temperature lamella of IRA 900 with visible clusters and voids. (E) Visible cluster stacking on IRA 900. (F) Resolved cluster stacking on IRA 900, showing stacks around 2 nm in scale.

Figures 6A and 6B show TEM images of the Fumasep FAA-3 film in the parallel (thickness) direction, revealing an orderly granular structure that corresponds to the diffraction results presented in Figure 2C. Figure 6C presents the TEM image of Fumasep FAA-3 in the perpendicular direction, further confirming the granular morphology observed in the AFM images (Figures 4A-C). Figure 6D shows the TEM image of an IRA 900 showing clusters and nano pores from the prepared lamella at room temperature. Figure 6E shows the cluster stacking with the resolved stacks around 2 nm in scale, which is shown in Figure 6F. These TEM results from IRA 900 reveal interesting features within the atomic scales, revealing a stacking nature in the clusters that we observed from X-ray scattering data (Figures 2B and 3D), AFM results (Figures 4D-F), and SEM images (Figures 5E and 5G). Moreover, the thickness of each layer in the clusters seems to be around 2 nm, where their stacking behavior could give more surface area during the MS DAC processes.

# 4 Conclusion

Based on the work presented, we have analyzed the structural features of commercially available AEM-DAC polymers, Fumasep, a membrane, and IRA 900, a macroporous resin, in detail using a combination of advanced microscopy and scattering techniques. X-ray diffraction and SAXS/WAXS analyses revealed molecular ordering and larger-scale structural organization within these materials, while humidity-dependent changes highlighted the impact of moisture on their structural properties. AFM surface analysis further validated these materials' nanoscale clustering, pores, and swelling at higher humidity. Notably, FIB-SEM lamella preparation and subsequent TEM analysis provided high-resolution structural insights into both Fumasep and IRA 900, revealing distinct granular and clustered features that align with previous scattering and surface analysis results.

The relationship between structure and function, and the structural changes in charged polymers involved in moisture swing direct air capture (MS DAC), are revealed by SAXS/WAXS studies at defined humidity levels from 25 to 95%, AFM studies submerged in water, and cryogenic FIB-SEM studies of the sample submerged in water. Swelling, shrinking, pores, and cluster stacking behavior might impact the moisture-driven DAC. Moderate swelling and shrinking during the MS process may benefit the DAC process, but excessive changes can compromise the long-term durability of these materials due to potential degradation. Additionally, recyclability and regeneration efficiency may decline. The pores that we see from AFM, FIB-SEM, and TEM are nanoporous in scale, which might facilitate the bulk water, ion, and gas transport during the MS-mediated DAC.

These findings not only deepen our understanding of the structural behavior of DAC materials but also underscore the importance of moisture-induced changes in the $CO_2$ capture and release processes. The knowledge gained from these studies lays the groundwork for the design of more reliable, sustainable, and efficient DAC polymers, enabling improvements in $CO_2$ capture technologies. Future work will focus on synthesizing and characterizing novel polymers based on these structural insights to optimize their performance in capturing $CO_2$ with minimal energy input.

**Acknowledgments**

The authors gratefully acknowledge financial support from the U.S. Department of Energy, Office of Science, Office of Basic Energy Sciences under Award Number DE-SC0023343, which supported the research and publication of this work and directly supported GY.

We acknowledge the use of facilities within the Eyring Materials Center at Arizona State University, supported in part by NNCI-ECCS-2025490, and would like to thank Anthony Woolson for his assistance with AFM method development, data collection, data reduction, and analysis, and Kenneth Mossman for his assistance with cryo-FIB-SEM method development, data collection, and analysis.

We also thank TESCAN USA Inc. for providing access to plasma FIB technology, which was essential for conducting the focused ion beam experiments described in this study. Additional support and instrumentation were provided by the Transmission Electron Microscopy (TEM) facility at the Singh Center for Nanotechnology, University of Pennsylvania. We also thank the Biodesign Center for Applied Structural Discovery at Arizona State University for support of this work.

We extend our sincere appreciation to the Mission DAC team, including Matthew Green (Co-Principal Investigator, Associate Professor, School for Engineering of Matter, Transport and Energy, Arizona State University), and Miguel Jose Yacaman (Co-Principal Investigator, Regents Professor, Applied Physics and Materials Science, Northern Arizona University) for their valuable contribution and support.

**Author contributions**

Gayathri Yogaganeshan (GY): Conceptualization, sample preparation, developing methodology, data collection, formal analysis, manuscript preparation
Rui Zhang: SAXS/WAXS method development, SAXS/WAXS data collection, SAXS/WAXS data analysis, reviewing and editing SAXS/WAXS methodology
Raimund Fromme: X-ray diffraction data collection
Sharang Sharang: Cryo FIB SEM data collection
Jamie Ford: Cryo-FIB-SEM, and TEM facility support
Douglas M Yates: TEM facility support
Marlene Velazco Medel: Research support, preparation of Figure 1
Martin Uher:   Cryo FIB SEM facility support
Justin Flory: Funding acquisition, project administration, review & editing
Jennifer Wade: Project PI, research investigation
Petra Fromme: Project co-PI, conceptualization, supervision, methodology development, finalizing the manuscript, corresponding author

**Data availability statement**

All data relevant to the study are included in the article.

**Supporting Information**

Supporting Information is available online or from the author.

**References**

1. Wade JL, Lopez Marques H, Wang W, Flory J, Freeman B. Moisture-driven CO2 pump for direct air capture. J Memb Sci 2023;685.
2. Lopez-Marques H, Wade JL, Sinyangwe SK, Guzzo S, Reimund KK, Smith L, et al. Moisture swing CO2 sorption in ion exchange resins for direct air capture.

3. Abdullatif Y, Sodiq A, Mir N, Bicer Y, Al-Ansari T, El-Naas MH, et al. Emerging trends in direct air capture of CO2: a review of technology options targeting net-zero emissions. RSC Adv 2023;13(9):5687–722.
4. Dutcher B, Fan M, Russell AG. Amine-based CO2 capture technology development from the beginning of 2013-A review. ACS Appl Mater Interfaces 2015;7(4):2137–48.
5. Mondal BK, Bandyopadhyay SS, Samanta AN. Kinetics of CO2 Absorption in Aqueous Hexamethylenediamine Blended N-Methyldiethanolamine. Ind Eng Chem Res 2017;56(50):14902–13.
6. Prajapati A, Sartape R, Rojas T, Dandu NK, Dhakal P, Thorat AS, et al. Migration-assisted, moisture gradient process for ultrafast, continuous CO2 capture from dilute sources at ambient conditions. Energy Environ Sci 2022;15(2):680–92.
7. Shi X, Xiao H, Azarabadi H, Song J, Wu X, Chen X, et al. Sorbenten zur direkten Gewinnung von CO 2 aus der Umgebungsluft. Angewandte Chemie 2020;132(18):7048–72.
8. Velazco-Medel MA, Niimoto KTM, Green MD. Exploring Phosphonium-Based Anion Exchange Polymers for Moisture Swing Direct Air Capture of Carbon Dioxide. Macromol Rapid Commun 2025;46(12).
9. Shi X, Li Q, Wang T, Lackner KS. Kinetic analysis of an anion exchange absorbent for CO2 capture from ambient air. PLoS One 2017;12(6).
10. Shi X, Xiao H, Kanamori K, Yonezu A, Lackner KS, Chen X. Moisture-Driven CO2 Sorbents. Joule 2020;4(8):1823–37.
11. Wang T, Lackner KS, Wright AB. Moisture-swing sorption for carbon dioxide capture from ambient air: A thermodynamic analysis. Physical Chemistry Chemical Physics 2013;15(2):504–14.
12. Wang T, Lackner KS, Wright A. Moisture swing sorbent for carbon dioxide capture from ambient air. Environ Sci Technol 2011;45(15):6670–5.
13. Lackner KS. Capture of carbon dioxide from ambient air. European Physical Journal: Special Topics 2009;176(1):93–106.
14. Zeman F. Energy and material balance of CO2 capture from ambient air. Environ Sci Technol 2007;41(21):7558–63.
15. Wang Y, Kim J, Marreiros J, Rangnekar N, Yuan Y, Johnson JR, et al. Investigation of Moisture Swing Adsorbents for Direct Air Capture by Dynamic Breakthrough Studies. ACS Sustain Chem Eng 2025;
16. Ziv N, Mondal AN, Weissbach T, Holdcroft S, Dekel DR. Effect of CO2 on the properties of anion exchange membranes for fuel cell applications. J Memb Sci 2019;586:140–50.
17. Pandey TP, Peters BD, Liberatore MW, Herring AM. Insight on Pure vs Air Exposed Hydroxide Ion Conductivity in an Anion Exchange Membrane for Fuel Cell Applications. ECS Trans 2014;64(3):1195–200.
18. Shi X, Xiao H, Lackner KS, Chen X. Capture CO 2 from Ambient Air Using Nanoconfined Ion Hydration . Angewandte Chemie 2016;128(12):4094–7.
19. Sarbaz S, Liu ZX, Feigenbaum H, Bayati S, Wang W, Wade J, et al. Characterizing and modeling the mechanical behavior of an anion exchange membrane for carbon capture applications. 2025;Available from: http://arxiv.org/abs/2508.01910
20. Barnes AM, Liu B, Buratto SK. Humidity-Dependent Surface Structure and Hydroxide Conductance of a Model Quaternary Ammonium Anion Exchange Membrane. Langmuir 2019;35(44):14188–93.

21. Colletta M, Yang Y, Goodge BH, Abruña HD, Kourkoutis LF. Overcoming Artifacts in Imaging Nanometer-thick Ionomer Layers in Anion Exchange Membrane Fuel Cells. Microscopy and Microanalysis 2022;28(S1):2210–2.
22. Tian J, Xie SH, Borucu U, Lei S, Zhang Y, Manners I. High-resolution cryo-electron microscopy structure of block copolymer nanofibres with a crystalline core. Nat Mater 2023;22(6):786–92.
23. Luo X, Rojas-Carbonell S, Yan Y, Kusoglu A. Structure-transport relationships of poly(aryl piperidinium) anion-exchange membranes: Eeffect of anions and hydration. J Memb Sci [Internet] 2020;598(November 2019):117680. Available from: https://doi.org/10.1016/j.memsci.2019.117680
24. Alexander S, Hellemans L, Marti O, Schneir J, Elings V, Hansma PK, et al. An atomic-resolution atomic-force microscope implemented using an optical lever. J Appl Phys 1989;65(1):164–7.
25. Najibah M, Kong J, Khalid H, Hnát J, Park HS, Bouzek K, et al. Pre-swelling of FAA3 membranes with water-based ethylene glycol solution to minimize dimensional changes after assembly into a water electrolyser: Effect on properties and performance. J Memb Sci 2023;670(November 2022).
26. Paajanen A, Zitting A, Rautkari L, Ketoja JA, Penttilä PA. Nanoscale Mechanism of Moisture-Induced Swelling in Wood Microfibril Bundles. Nano Lett 2022;22(13):5143–50.
27. Zheng Y, Ash U, Pandey RP, Ozioko AG, Ponce-González J, Handl M, et al. Water Uptake Study of Anion Exchange Membranes. Macromolecules 2018;51(9):3264–78.
28. Pandey TP, Maes AM, Sarode HN, Peters BD, Lavina S, Vezzù K, et al. Interplay between water uptake, ion interactions, and conductivity in an e-beam grafted poly(ethylene-co-tetrafluoroethylene) anion exchange membrane. Physical Chemistry Chemical Physics 2015;17(6):4367–78.
29. Hatakenaka S, Takata H, Mizuno N, Mamiya Y, Nishikawa M, Fukada S, et al. Adsorption of water vapor on a polymer electrolyte membrane. Int J Hydrogen Energy 2008;33(13):3368–72.
30. Mangiagli PM, Ewing CS, Xu K, Wang Q, Hickner MA. Dynamic water uptake of flexible polymer networks ion-containing polymer networks. In: Fuel Cells. 2009. page 432–8.
31. Milani M, Drobne D, Tatti F. How to study biological samples by FIB/SEM? … and Educational Topics in Microscopy A … [Internet] 2007;(January):787–94. Available from: http://www.formatex.org/microscopy3/pdf/pp787-794.pdf
32. Lešer V, Drobne D, Pipan Ž, Milani M, Tatti F. Comparison of different preparation methods of biological samples for FIB milling and SEM investigation. J Microsc 2009;233(2):309–19.

# Comprehensive Structural Characterization of Charged Polymers Involved in Moisture-Driven Direct Air Capture

Gayathri Yogaganeshan, Rui Zhang, Raimund Fromme, Sharang Sharang, Jamie Ford, Douglas M Yates, Marlene Velazco Medel, Martin Uher, Justin Flory, Jennifer Wade, Petra Fromme[*]

## Supporting information

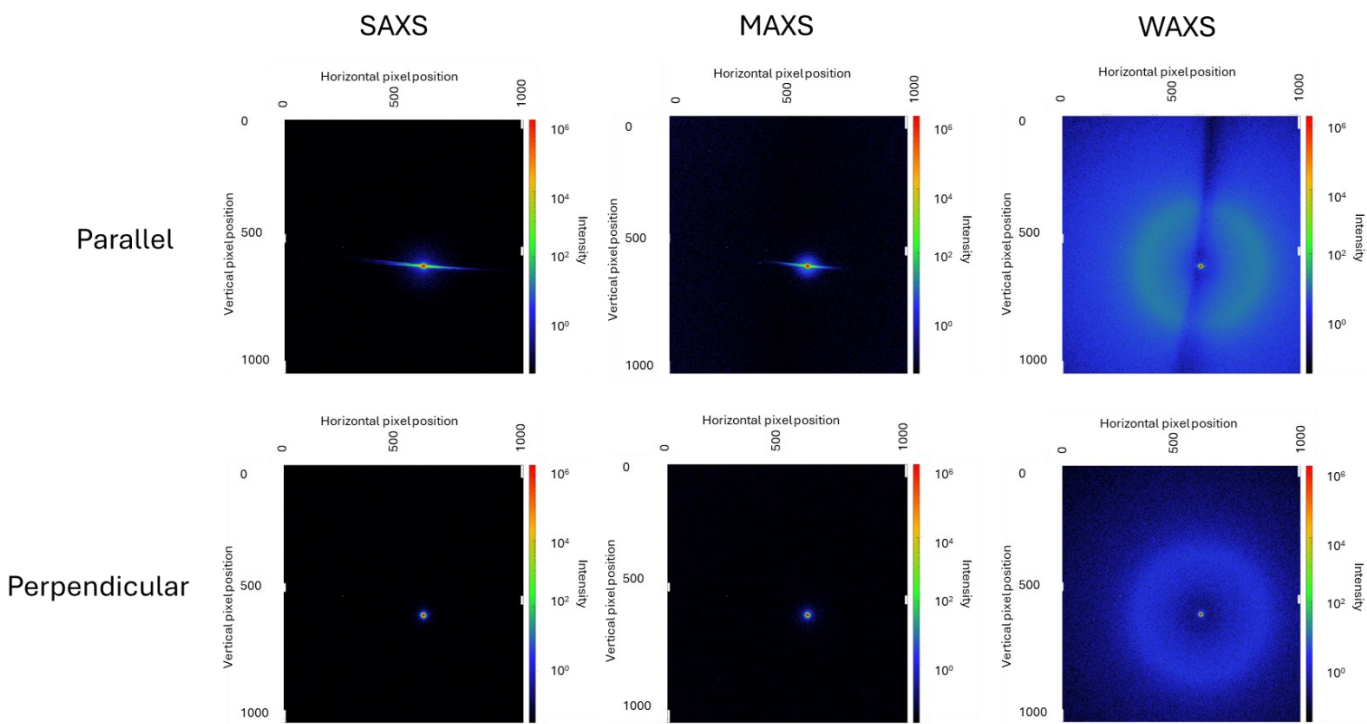

**Figure SI1**: SAXS, MAXS, and WAXS scattering patterns of Fumasep in parallel and perpendicular direction at ambient (~30% RH) conditions.

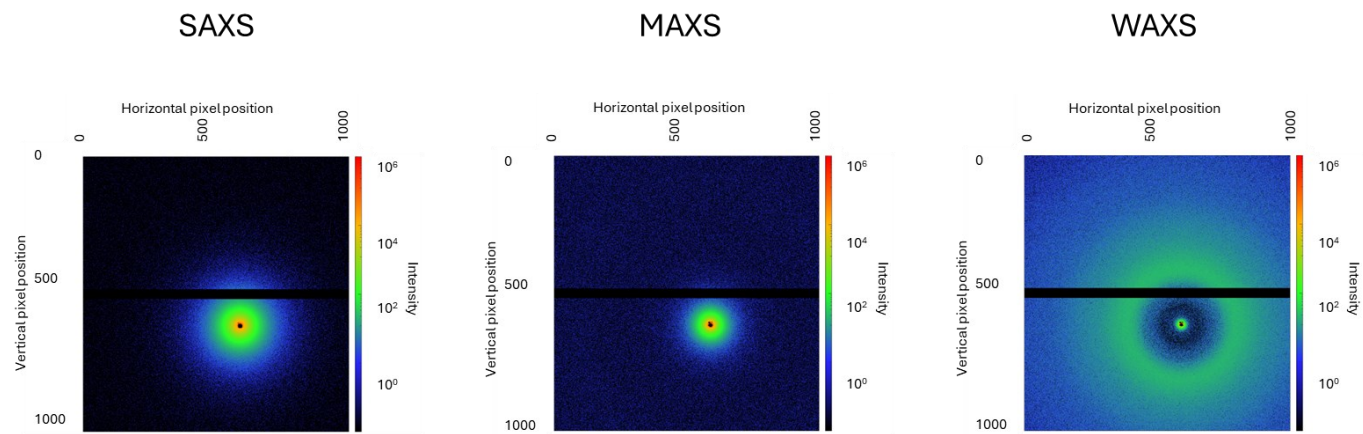

**Figure SI2**: SAXS, MAXS, and WAXS scattering patterns of IRA 900 at ambient (~30% RH) conditions.

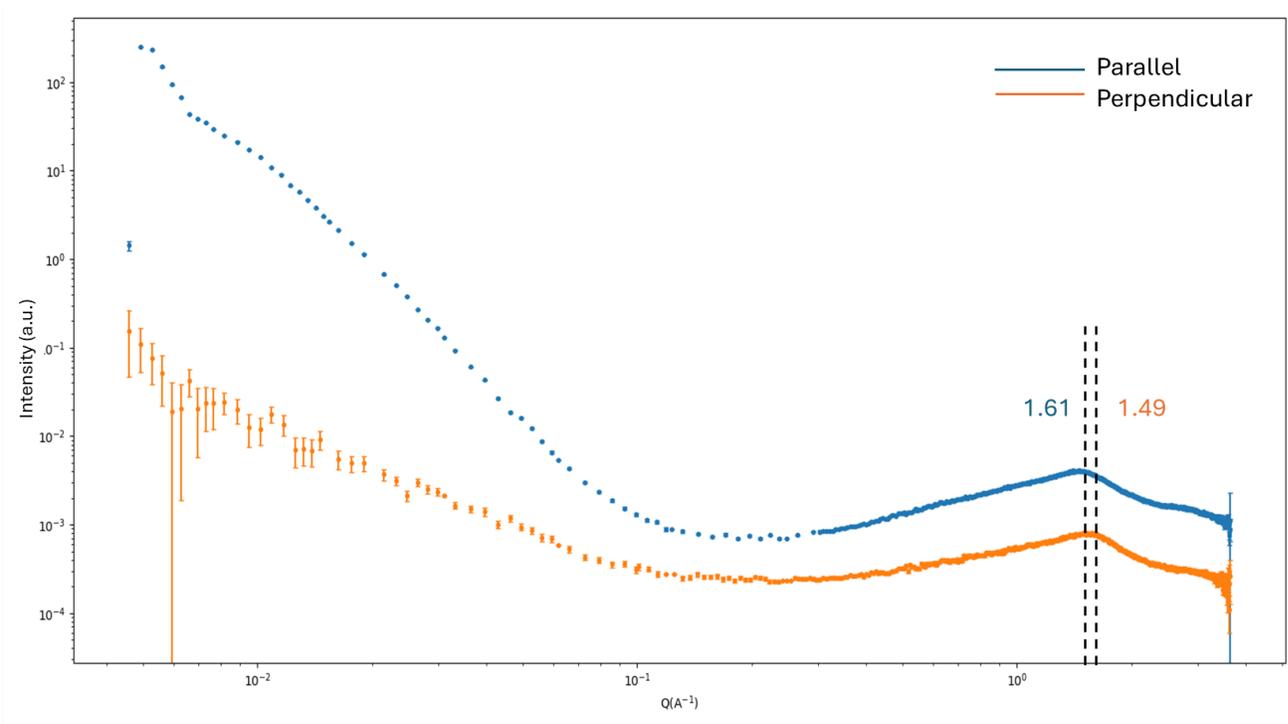

**Figure SI3**: Correlating the scattering vector to real length scales from intensity (a.u.) vs scattering vector Q (Å$^{-1}$) plot of Fumasep at ambient (30%) RH conditions in parallel and perpendicular directions.

Using $Q = \frac{2\pi}{D}$, where $\pi = 3.14$, eventually $D = \frac{2\pi}{Q}$

WAXS peak in the parallel direction = $D = \frac{2*3.14}{1.61 \text{ Å}^{-1}} = 3.90$ Å

WAXS peak in the perpendicular direction = $D = \frac{2*3.14}{1.49 \text{ Å}^{-1}} = 4.21$ Å

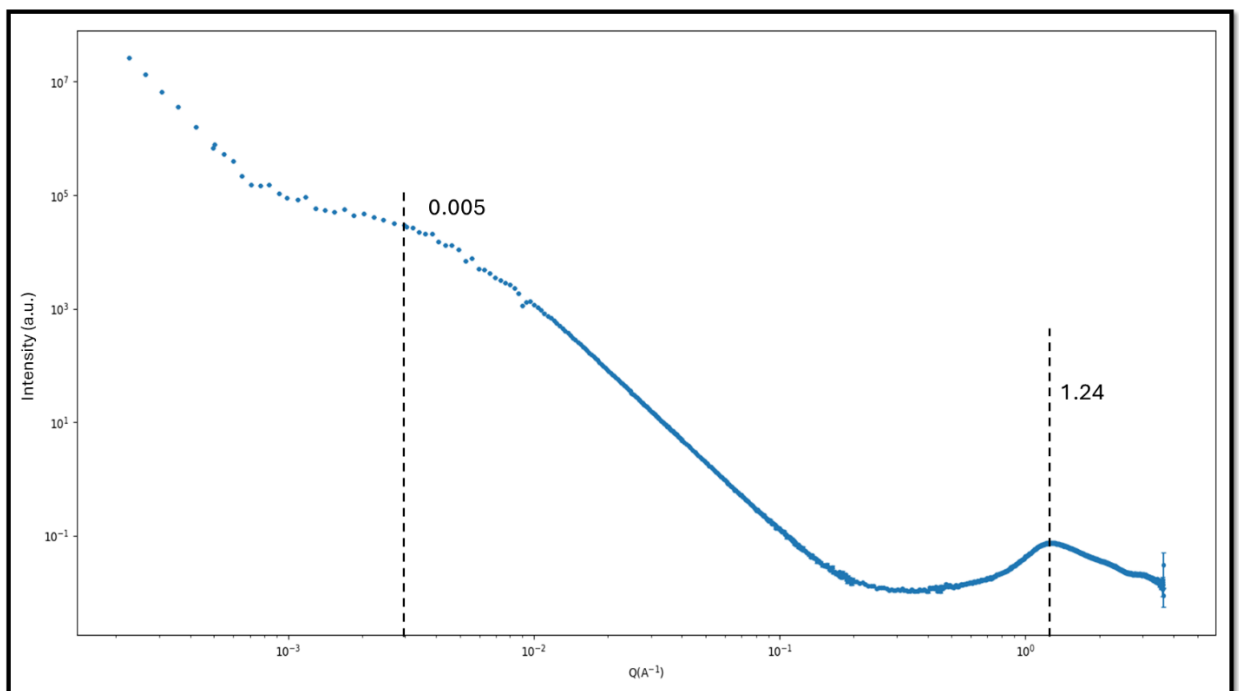

**Figure SI4**: Correlating the scattering vector to real length scales from intensity (a.u.) vs scattering vector Q (Å$^{-1}$) plot of IRA 900 at ambient (30%) RH conditions.

Using $Q = \frac{2\pi}{D}$, where $\pi = 3.14$, eventually $D = \frac{2\pi}{Q}$

WAXS peak = $D = \frac{2*3.14}{1.24 \text{ Å}^{-1}} = 5.06$ Å

SAXS plateau = $D = \frac{2*3.14}{0.005 \text{ Å}^{-1}} = 1256$ Å

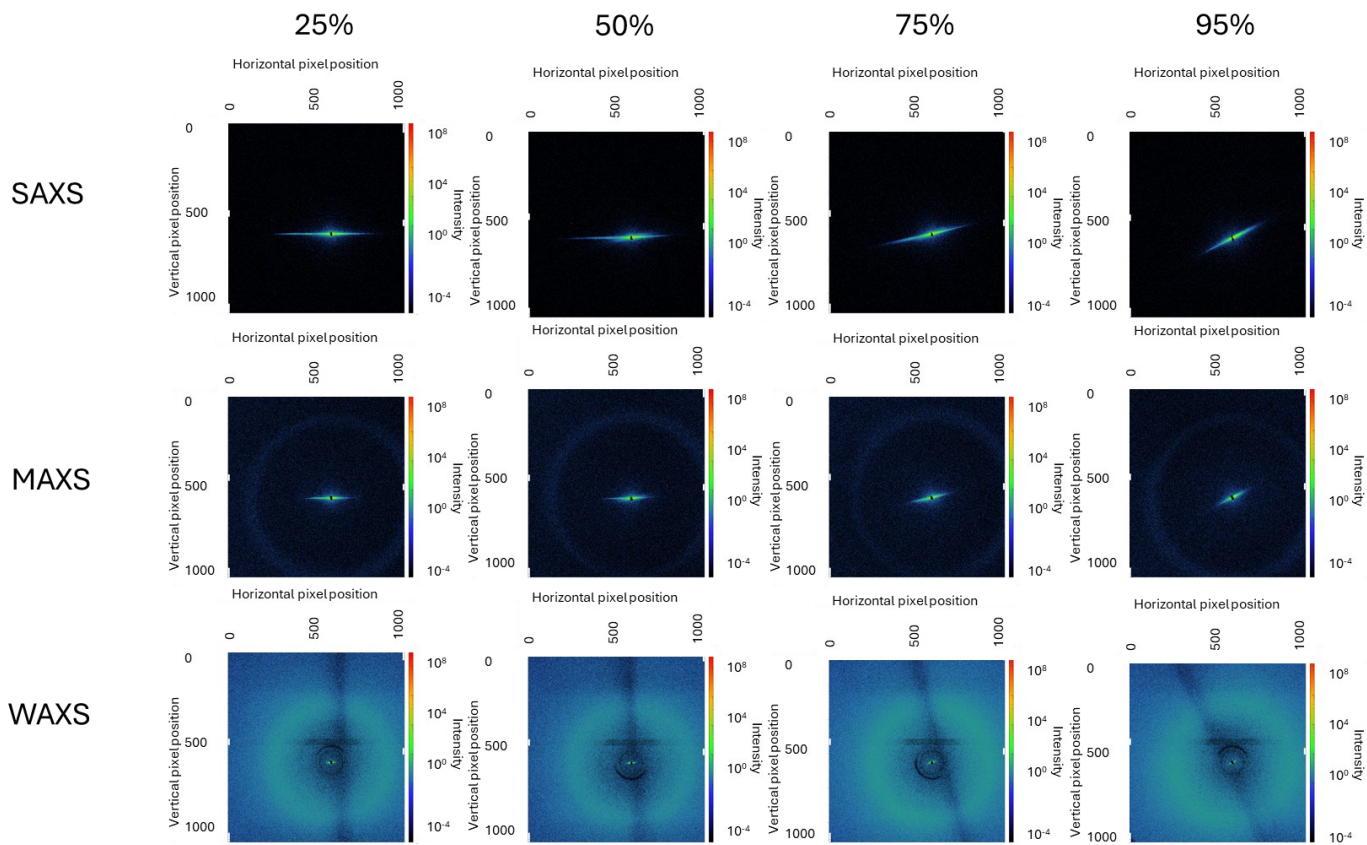

**Figure SI5**: SAXS, MAXS, and WAXS scattering patterns of Fumasep in the parallel direction at 25%, 50%, 75%, and 95% RH conditions.

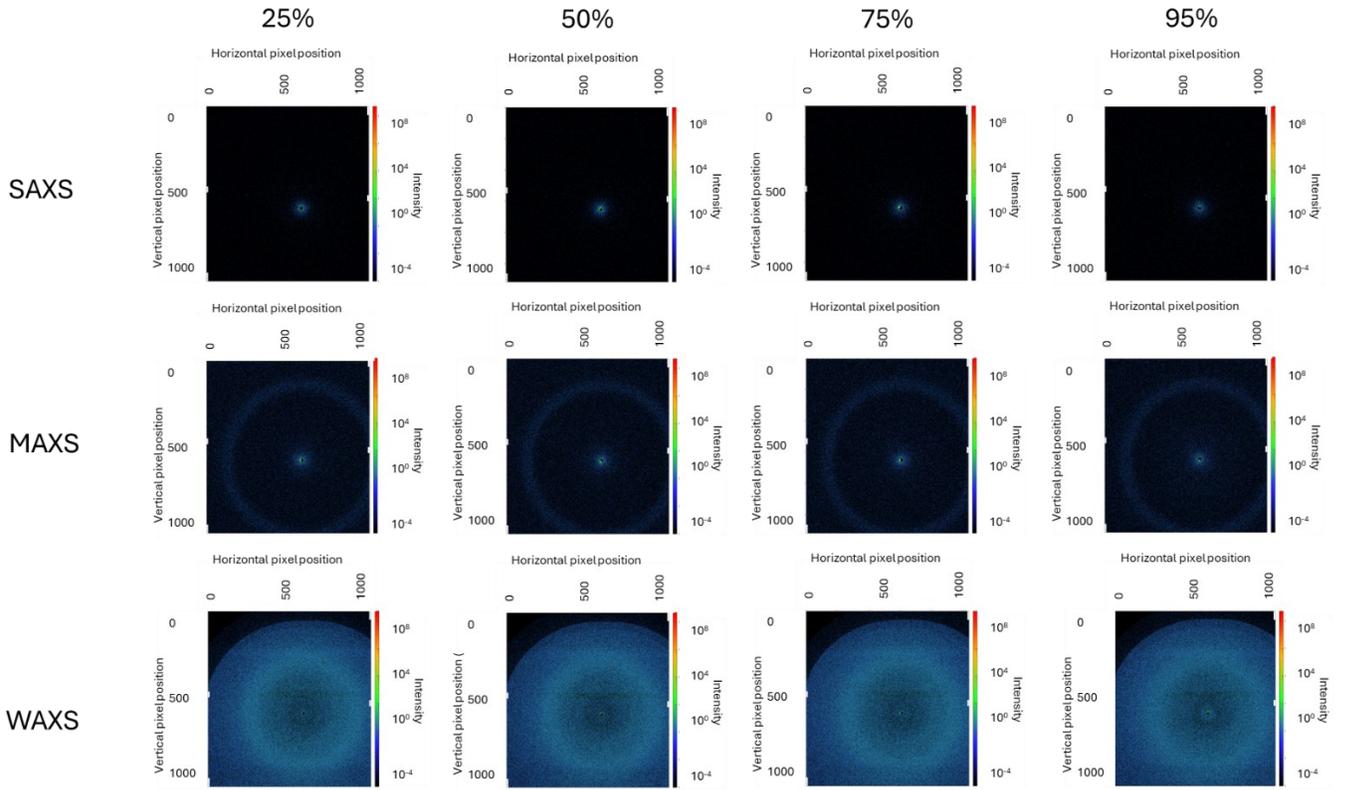

**Figure SI6**: SAXS, MAXS, and WAXS scattering patterns of Fumasep in the perpendicular direction at 25%, 50%, 75%, and 95% RH conditions.

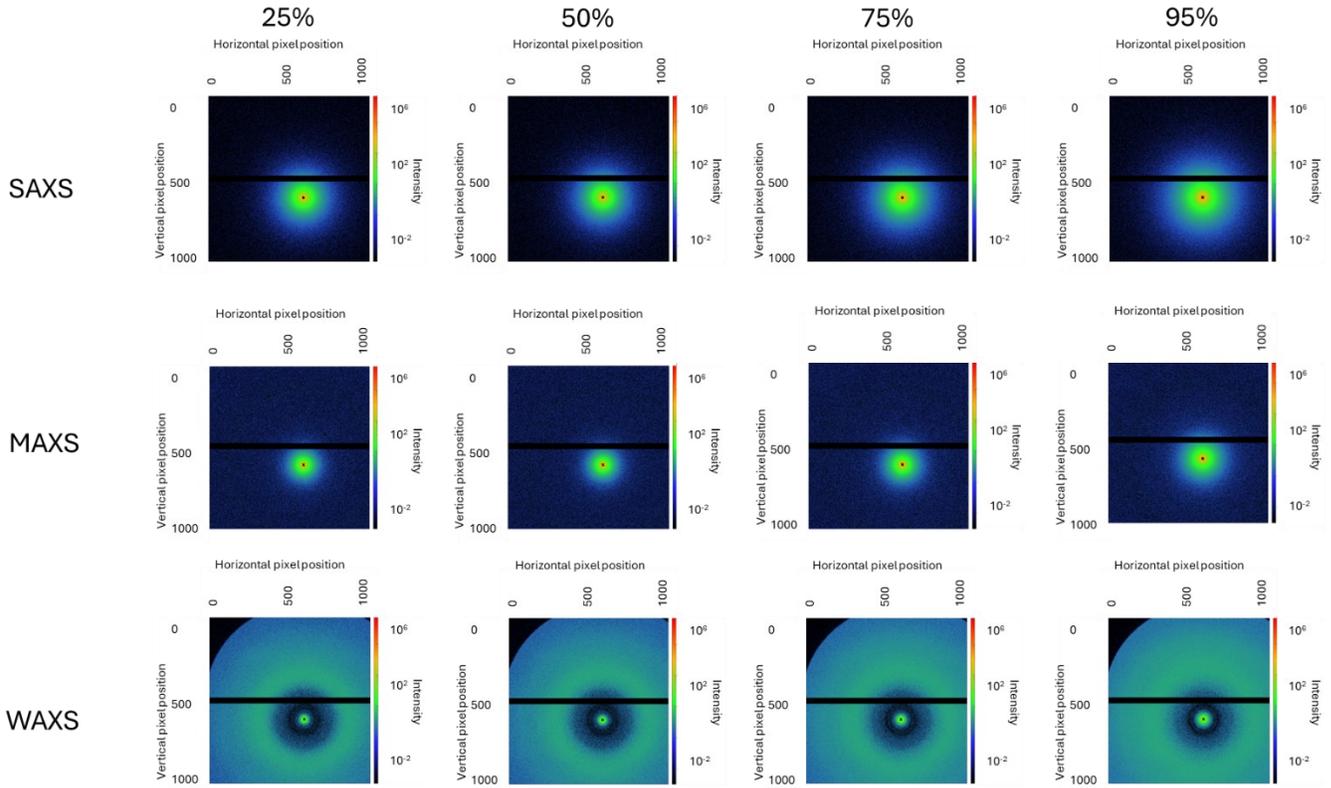

**Figure SI7**: SAXS, MAXS, and WAXS scattering patterns of IRA 900 at 25%, 50%, 75%, and 95% RH conditions.

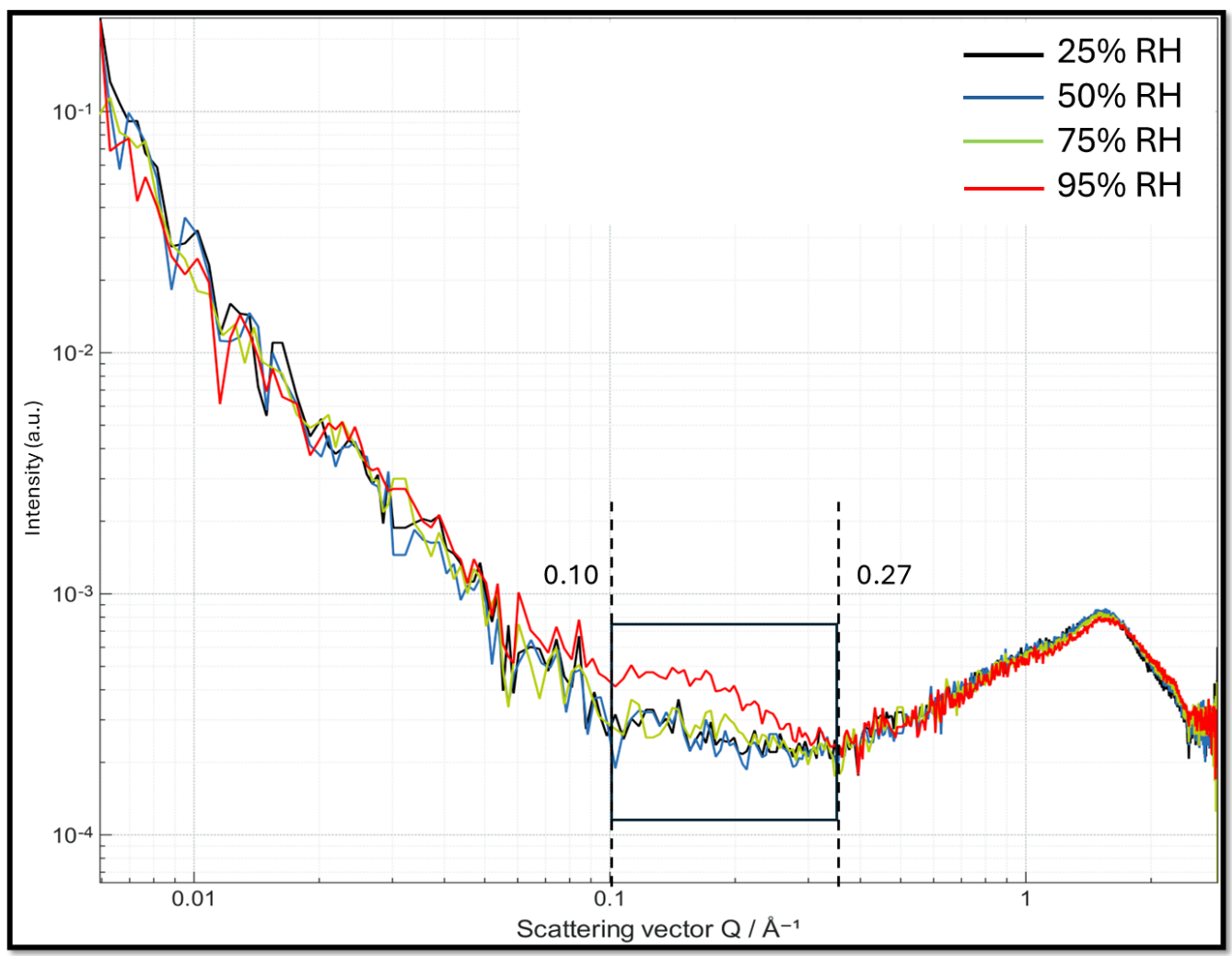

**Figure SI8**: Correlating the scattering vector to real length scales of the hump observed at 95% RH (red plot) on Fumasep in perpendicular direction in the SAXS region from intensity (a.u.) vs scattering vector Q (Å$^{-1}$) plot.

Using $Q = \frac{2\pi}{D}$, where $\pi = 3.14$, eventually $D = \frac{2\pi}{Q}$

SAXS hump at 95% (red plot) =
$D = \frac{2*3.14}{0.27 \text{ Å}^{-1}} = 23.26 \text{ Å}$
To
$D = \frac{2*3.14}{0.10 \text{ Å}^{-1}} = 62.8 \text{ Å}$

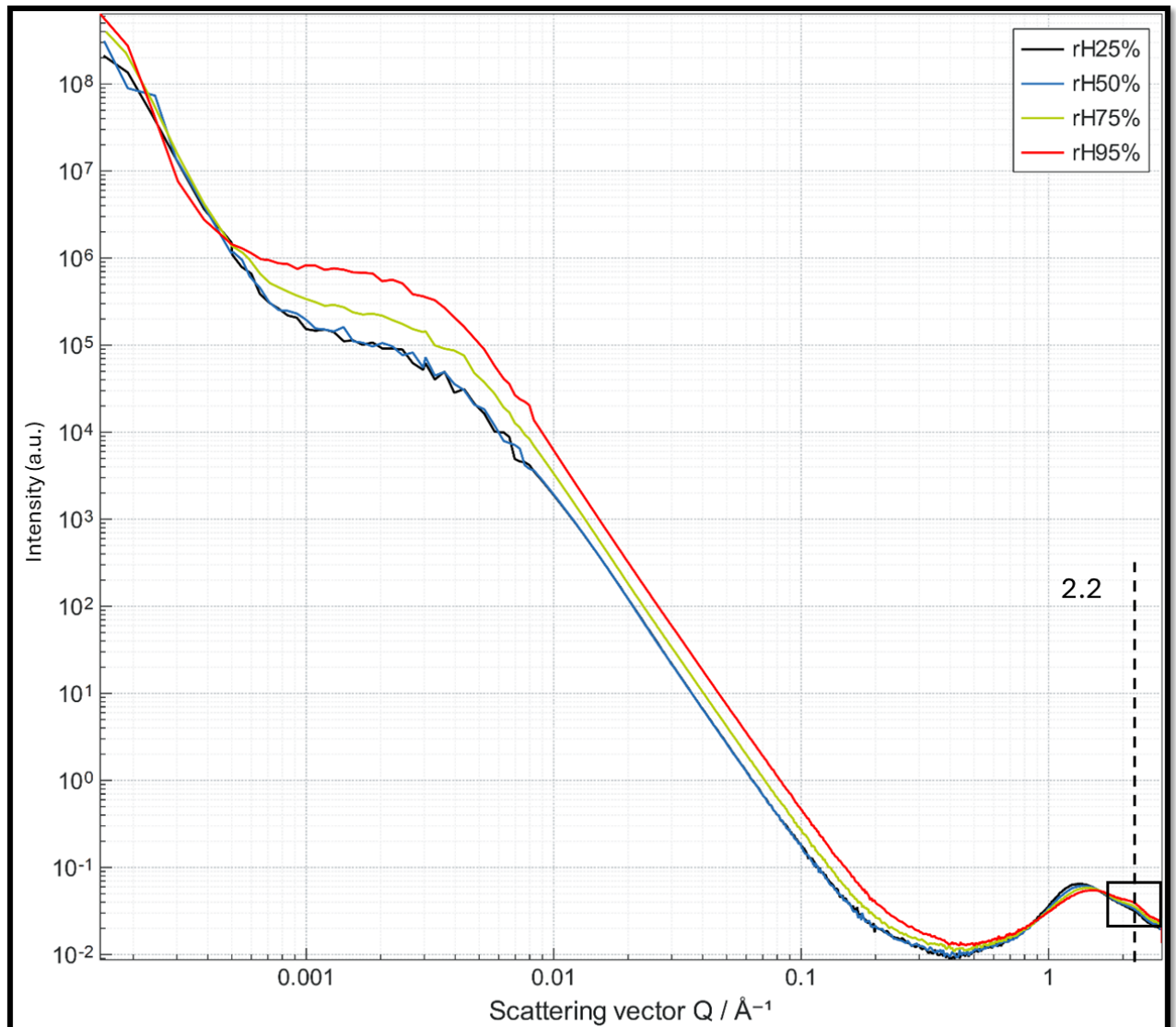

**Figure SI9**: Correlating the scattering vector to real length scales of the additional prominent small hump observed in IRA 900 at increasing RH from intensity (a.u.) vs scattering vector Q (Å$^{-1}$) plot.

Using $Q = \frac{2\pi}{D}$, where $\pi = 3.14$, eventually $D = \frac{2\pi}{Q}$

WAXS hump = $D = \frac{2*3.14}{2.2 \text{ Å}^{-1}} = 2.85$ Å